\documentclass[twocolumn,twocolappendix]{aastex631}
\usepackage{amsmath,amssymb,mathtools,etoolbox}
\usepackage{booktabs}
\usepackage{array}
\usepackage{makecell}
\usepackage{multirow,stackengine}            
\usepackage{capt-of}
\usepackage{tabularx,colortbl}
\setstackEOL{\cr}
\usepackage{xcolor}

\newcommand{\msun}{$ M_\odot$}

\newcommand{\mjybeam}{$\mathrm{mJy\,beam^{-1}}$}

\newcommand{\parcsec}{\mbox{$.\!\!\arcsec$}}
\newcommand{\parcmin}{\mbox{$.\!\!\arcmin$}}

\shorttitle{Dense Cores Are Scarce in High-latitude Clouds} \shortauthors{Xu et al.}
\begin{document}

\title{On the Scarcity of Dense Cores ($n>10^{5}$\,cm$^{-3}$) in High-latitude Planck Galactic Cold Clumps}

\correspondingauthor{Ke Wang}
\email{kwang.astro@pku.edu.cn}

\AuthorCollaborationLimit=50

\author[0000-0001-5950-1932]{Fengwei Xu}
\affiliation{Kavli Institute for Astronomy and Astrophysics, Peking University, Beijing 100871, People’s Republic of China}
\affiliation{Department of Astronomy, School of Physics, Peking University, Beijing, 100871, People's Republic of China}
% fengwei.astro@pku.edu.cn

\author[0000-0002-7237-3856]{Ke Wang}
\affiliation{Kavli Institute for Astronomy and Astrophysics, Peking University, Beijing 100871, People’s Republic of China}
% kwang.astro@pku.edu.cn

\author[0000-0002-5286-2564]{Tie Liu}
\affiliation{Shanghai Astronomical Observatory, Chinese Academy of Sciences, 80 Nandan Road, Shanghai 200030, People's Republic of China}
% liutie@shao.ac.cn

\author[0000-0002-5881-3229]{David Eden}
\affiliation{Armagh Observatory and Planetarium, College Hill, Armagh, BT61 9DG, UK}
% david.eden@armagh.ac.uk

\author[0000-0001-8315-4248]{Xunchuan Liu}
\affiliation{Shanghai Astronomical Observatory, Chinese Academy of Sciences, 80 Nandan Road, Shanghai 200030, People's Republic of China}
% liuxunchuan@qq.com

\author[0000-0002-5809-4834]{Mika Juvela}
\affiliation{Department of Physics, University of Helsinki, PO Box 64, FI-00014 Helsinki, Finland}
% mika.juvela@helsinki.fi

\author[0000-0002-3938-4393]{Jinhua He}
\affiliation{Yunnan Observatories, Chinese Academy of Sciences, 396 Yangfangwang,
Guandu District, Kunming, 650216, People's Republic of China}
\affiliation{Chinese Academy of Sciences South America Center for Astronomy,National
Astronomical Observatories, CAS, Beijing 100101, People's Republic of China}
\affiliation{Departamento de Astronom\'{\i}a, Universidad de Chile, Las Condes, 7591245 Santiago, Chile}
% jinhuahe2009@gmail.com

\author[0000-0002-6773-459X]{Doug Johnstone}
\affiliation{NRC Herzberg Astronomy and Astrophysics, 5071 West Saanich Road, Victoria, BC, V9E 2E7, Canada}
\affiliation{Department of Physics and Astronomy, University of Victoria, 3800 Finnerty Road, Elliot Building, Victoria, BC, V8P 5C2, Canada}
% doug.johnstone@gmail.com

\author[0000-0002-6622-8396]{Paul Goldsmith}
\affiliation{Jet Propulsion Laboratory, California Institute of Technology, Pasadena CA 91109, USA}
% paul.f.goldsmith@jpl.nasa.gov

\author[0000-0003-1649-7958]{Guido Garay}
\affiliation{Departamento de Astronom\'{\i}a, Universidad de Chile, Las Condes, 7591245 Santiago, Chile}
% guido@das.uchile.cl

\author[0000-0002-5076-7520]{Yuefang Wu}
\affiliation{Department of Astronomy, School of Physics, Peking University, Beijing, 100871, People's Republic of China}
\affiliation{Kavli Institute for Astronomy and Astrophysics, Peking University, Beijing 100871, People’s Republic of China}
% ywu@pku.edu.cn

\author[0000-0002-6386-2906]{Archana Soam}
\affiliation{Indian Institute of Astrophysics, II Block, Koramangala, Bengaluru 560034, India}
% archanasoam.bhu@gmail.com

\author[0000-0003-1665-6402]{Alessio Traficante}
\affiliation{IAPS-INAF, Via Fosso del Cavaliere, 100, I-00133 Rome, Italy}
% alessio.traficante@inaf.it

\author{Isabelle Ristorcelli}
\affiliation{3 Université de Toulouse, UPS-OMP, IRAP, F-31028 Toulouse cedex 4, France}
% isabelle.ristorcelli@irap.omp.eu

\author[0000-0003-0693-2477]{Edith Falgarone}
\affiliation{LPENS, Ecole Normale Supérieure, Université PSL, CNRS, Sorbonne Université, Université de Paris, 75005 Paris, France}
% edith.falgarone@lra.ens.fr

\author[0000-0002-9774-1846]{Huei-Ru Vivien Chen}
\affiliation{Institute of Astronomy and Department of Physics, National Tsing Hua University, Hsinchu 30013, Taiwan}
% hchen@phys.nthu.edu.tw

\author[0000-0001-9304-7884]{Naomi Hirano}
\affiliation{Institute of Astronomy and Astrophysics, Academia Sinica, No.1, Sec. 4, Roosevelt Road, Taipei 10617, Taiwan}
% hirano@asiaa.sinica.edu.tw

\author[0000-0001-8746-6548]{Yasuo Doi}
\affiliation{Department of Earth Science and Astronomy, Graduate School of Arts and Sciences, The University of Tokyo, 3-8-1 Komaba, Meguro, Tokyo 153-8902, Japan}
% doiutokyo@gmail.com

\author[0000-0003-4022-4132]{Woojin Kwon}
\affiliation{Department of Earth Science Education, Seoul National University, 1 Gwanak-ro, Gwanak-gu, Seoul 08826, Republic of Korea}
\affiliation{SNU Astronomy Research Center, Seoul National University, 1 Gwanak-ro, Gwanak-gu, Seoul 08826, Republic of Korea}
% wkwon@snu.ac.kr

\author[0000-0002-7126-691X]{Glenn J. White}
\affiliation{School of Physical Sciences, The Open University, Walton Hall, Milton Keynes, MK7 6AA, UK}
\affiliation{RAL Space, STFC Rutherford Appleton Laboratory, Chilton, Didcot, Oxfordshire, OX11 0QX, UK}
% Glenn.White@stfc.ac.uk

\author[0000-0002-1178-5486]{Anthony Whitworth}
\affiliation{School of Physics and Astronomy, Cardiff University, The Parade, Cardiff, CF24 3AA, UK}
% Anthony.Whitworth@astro.cf.ac.uk

\author[0000-0002-7125-7685]{Patricio Sanhueza}
\affiliation{National Astronomical Observatory of Japan, National Institutes of Natural Sciences, 2-21-1 Osawa, Mitaka, Tokyo 181-8588, Japan}
\affiliation{Astronomical Science Program, The Graduate University for Advanced Studies, SOKENDAI, 2-21-1 Osawa, Mitaka, Tokyo 181-8588, Japan}
% patosanhueza@gmail.com

\author[0000-0002-6529-202X]{Mark G. Rawlings}
\affiliation{Gemini Observatory/NSF’s NOIRLab, 670 N. A‘ohōkū Place, Hilo, HI 96720, USA}
\affiliation{East Asian Observatory, 660 N. A‘ohōkū Place, University Park, Hilo, HI 96720, USA}
% mark.rawlings@noirlab.edu

\author[0000-0001-5403-356X]{Dana Alina}
\affiliation{Nazarbayev University, Kabanbay Batyr Ave, 53, Astana 010000 Kazakhstan}
% dana.alina@nu.edu.kz

\author[0000-0003-4659-1742]{Zhiyuan Ren}
\affiliation{National Astronomical Observatories, Chinese Academy of Sciences, Datun Rd. A20, Beijing, People's Republic of China}
% renzy@nao.cas.cn

\author[0000-0002-3179-6334]{Chang Won Lee}
\affiliation{Korea Astronomy and Space Science Institute, 776 Daedeokdae-ro, Yuseong-gu, Daejeon 34055, Republic of Korea}
\affiliation{University of Science and Technology, Korea (UST), 217 Gajeong-ro, Yuseong-gu, Daejeon 34113, Republic of Korea}
% cwl@kasi.re.kr

\author[0000-0002-8149-8546]{Ken'ichi Tatematsu}
\affiliation{Nobeyama Radio Observatory, National Astronomical Observatory of Japan, National Institutes of Natural Sciences, 462-2 Nobeyama, Minamimaki, Minamisaku, Nagano 384-1305, Japan}
\affiliation{Astronomical Science Program, The Graduate University for Advanced Studies, SOKENDAI, 2-21-1 Osawa, Mitaka, Tokyo 181-8588, Japan}
% k.tatematsu@nao.ac.jp

\author[0000-0002-4428-3183]{Chuan-Peng Zhang}
\affiliation{National Astronomical Observatories, Chinese Academy of Sciences, Beijing 100101, China}
\affiliation{Guizhou Radio Astronomical Observatory, Guizhou University, Guiyang 550000, China}
% cpzhang@nao.cas.cn

\author[0000-0003-0356-818X]{Jianjun Zhou}
\affiliation{XingJiang Astronomical Observatory, Chinese Academy of Sciences, Urumqi 830011, PR China}
\affiliation{Key Laboratory of Radio Astronomy, Chinese Academy of Sciences, Urumqi 830011, PR China}
\affiliation{Xinjiang Key Laboratory of Radio Astrophysics, Urumqi 830011, PR China}
% zhoujj@xao.ac.cn

\author[0000-0001-5522-486X]{Shih-Ping Lai}
\affiliation{Institute of Astronomy and Astrophysics, Academia Sinica, No.1, Sec. 4, Roosevelt Road, Taipei 10617, Taiwan}
% slai@gapp.nthu.edu.tw

\author[0000-0003-1140-2761]{Derek Ward-Thompson}
\affiliation{Jeremiah Horrocks Institute, University of Central Lancashire, Preston PR1 2HE, UK}
% DWard-Thompson@uclan.ac.uk

\author[0000-0003-4603-7119]{Sheng-Yuan Liu}
\affiliation{Institute of Astronomy and Astrophysics, Academia Sinica, No.1, Sec. 4, Roosevelt Road, Taipei 10617, Taiwan}
% syliu@asiaa.sinica.edu.tw

\author[0000-0002-2826-1902]{Qilao Gu}
\affiliation{Shanghai Astronomical Observatory, Chinese Academy of Sciences, 80 Nandan Road, Shanghai 200030, People's Republic of China}
% qlgu@shao.ac.cn

\author[0000-0003-4761-6139]{Eswaraiah Chakali}
\affiliation{Indian Institute of Science Education and Research (IISER) Tirupati, Rami Reddy Nagar, Karakambadi Road, Mangalam (P.O.), Tirupati 517 507, India}
% eswaraiahc@labs.iisertirupati.ac.in

\author{Lei Zhu}
\affiliation{National Astronomical Observatories, Chinese Academy of Sciences, Datun Rd. A20, Beijing, People's Republic of China}
% lzhupku@gmail.com

\author[0000-0002-5065-9175]{Diego Mardones}
\affiliation{Departamento de Astronom\'{\i}a, Universidad de Chile, Las Condes, 7591245 Santiago, Chile}

\author[0000-0002-5310-4212]{L. Viktor T\'oth}
\affiliation{Institute of Physics and Astronomy, E\"otv\"os Lor\`and University, P\'azm\'any P\'eter s\'et\'any 1/A, H-1117 Budapest, Hungary}

\begin{abstract}

High-latitude ($|b|>30\arcdeg$) molecular clouds have virial parameters that exceed 1, but whether these clouds can form stars has not been studied systematically. Using JCMT SCUBA-2 archival data, we surveyed 70 fields that target high-latitude Planck Galactic cold clumps (HLPCs) to find dense cores with density of $10^{5}$--$10^{6}$\,cm$^{-3}$ and size of $<0.1$\,pc. The sample benefits from both the representativeness of the parent sample and its coverage of the densest clumps at the high column density end ($>1\times10^{21}$\,cm$^{-2}$). At an average rms of 15\,\mjybeam, we detected Galactic dense cores in only one field, G6.04+36.77 (L183) while also identifying 12 extragalactic objects and two young stellar objects. Compared to the low-latitude clumps, dense cores are scarce in HLPCs. With synthetic observations, the densities of cores are constrained to be $n_c\lesssim10^5$\,cm$^{-3}$ should they exist in HLPCs. Low-latitude clumps, Taurus clumps, and HLPCs form a sequence where a higher virial parameter corresponds to a lower dense-core detection rate. If HLPCs were affected by the Local Bubble, the scarcity should favor turbulence-inhibited rather than supernova-driven star formation. Studies of the formation mechanism of the L183 molecular cloud are warranted. 

\end{abstract}


%%%%%%%%%%%%%%%%%%%%%%%%%%%%%%%%%%%%%%%%%%
%%%%%%%%%%%% Introduction %%%%%%%%%%%%%%%%
%%%%%%%%%%%%%%%%%%%%%%%%%%%%%%%%%%%%%%%%%%

\section{Introduction} \label{sec:intro}

The high latitude (HL) of the Milky Way, also called the ``underwater iceberg'' guards its secrets about molecular gas and star formation, due in part to the limited scope of previous CO surveys \citep{Xu2021HGaL}. Observational challenges essentially originate from the large area of the HL, leading to much longer integration times compared to those required for blind surveys of the Galactic plane. The \textit{Planck} satellite provides an unprecedented all-sky census of the coldest \citep[6--20\,K with a median value of $\sim$14\,K;][]{Planck2011XXIII-Properties} Galactic objects by combining the highest-frequency channels of the \textit{Planck} survey 353--857\,GHz (i.e., 850--350\,$\mu$m) with the far-infrared IRAS 100\,$\mu$m data \citep{IRAS1984,IRIS2005}. As a result, the \textit{Planck} team has cataloged 13,188 \textit{Planck} Galactic Cold Clumps \citep[PGCCs;][]{Planck2016XXVIII-PGCCs}, including 793 with absolute value of Galactic latitude higher than 30$^{\circ}$, a group we refer to as high-latitude Planck cold clumps (HLPCs). Benefiting from the unbiased nature of its parent sample, the 793 HLPCs are the least-biased sample of HL cold dust clumps, therefore serving as a foundation for studying the properties of the HL molecular gas and investigating the initial condition of star formation \citep{Wu2012PGCCs,Liu2013CO}. 

Our previous work performed a $^{12}$CO/$^{13}$CO/C$^{18}$O\,(1-0) survey toward 41 early cold cores \citep[ECCs; i.e., most reliable detections of PGCCs with signal-to-noise ratio $>15$;][]{Planck2011VII-ECC} with the Purple Mountain Observatory (PMO) 13.7\,m millimeter-wavelength telescope \citep{Xu2021HGaL}. Although detected CO cores have a typical density of several times $10^{4}$\,cm$^{-3}$, consistent with what has been found in nearby molecular cloud cores \citep{Benson1983HC5N-Cores,Myers1983CO-Cores,Myers1983NH3-Cores,Myers1983Subsonic-Cores,Benson1989Cores}, the turbulent energy is significantly higher than the gravitational energy, with median virial parameters of $\sim35$ \citep{Xu2021HGaL}. Therefore, our CO surveys unveiled a highly turbulent, diffuse molecular gas environment as a first glimpse of the initial conditions of star formation in the HL clouds. 

Stars form in dense cores \citep{Shu1987SF} with a typical size of $\lesssim0.1$\,pc and density of $10^{5}$--$10^{6}$\,cm$^{-3}$ \citep{WT1994JCMT,WT1999IRAM30m,Kirk2005JCMT}. The CO\,(1-0) transitions suffer from optical thickness and depletion at low temperature, so it is hard to probe the densest regions of molecular clouds. For example, a high-latitude cloud L1780 shows a cometary morphology and a CO core \citep{Toth1995L1780} but contains no dense core in our survey (field G358.96+36.81). Furthermore, the angular resolution of \textit{Planck} used for the extraction of PGCCs is $\sim5$\arcmin, corresponding to 0.3\,pc at a typical distance of HL clouds of 200\,pc \citep{Xu2021HGaL}, which is marginal for resolving dense cores. Working at 850\,$\mu$m with an effective beam FWHM of 14\parcsec6, Submillimetre Common-User Bolometer Array 2 \citep[SCUBA-2;][]{Dempsey2013SCUBA-2} provides $\sim20$ times better resolution than the \textit{Planck}, pinpointing cold dense cores inside the molecular clouds.

In this Letter, we perform a systematic search for dense cores within 70 HLPCs using the latest JCMT SCUBA-2 archival data. The sample selection and distance estimation are summarized in Section\,\ref{sec:data}. As shown in Section\,\ref{sec:results}, dense cores are only identified in one HLPC (G6.04+36.77). In Section\,\ref{sec:discuss}, we show the robustness of the scarcity of Galactic dense cores in HLPCs, investigate the upper limit of the dense core density, and then discuss star formation picture at high latitude. Finally, we give a brief summary in Section\,\ref{sec:conclude}. 

%%%%%%%%%%%%%%%%%%%%%%%%%%%%%%%%%%%%%%%%%%
%%%%%%%%%%%%%%%%% Data %%%%%%%%%%%%%%%%%%%
%%%%%%%%%%%%%%%%%%%%%%%%%%%%%%%%%%%%%%%%%%

\section{Data}
\label{sec:data}

\subsection{Sample Selection}
\label{data:selection}

\begin{figure*}[!t]
\includegraphics[width=1.0\linewidth]{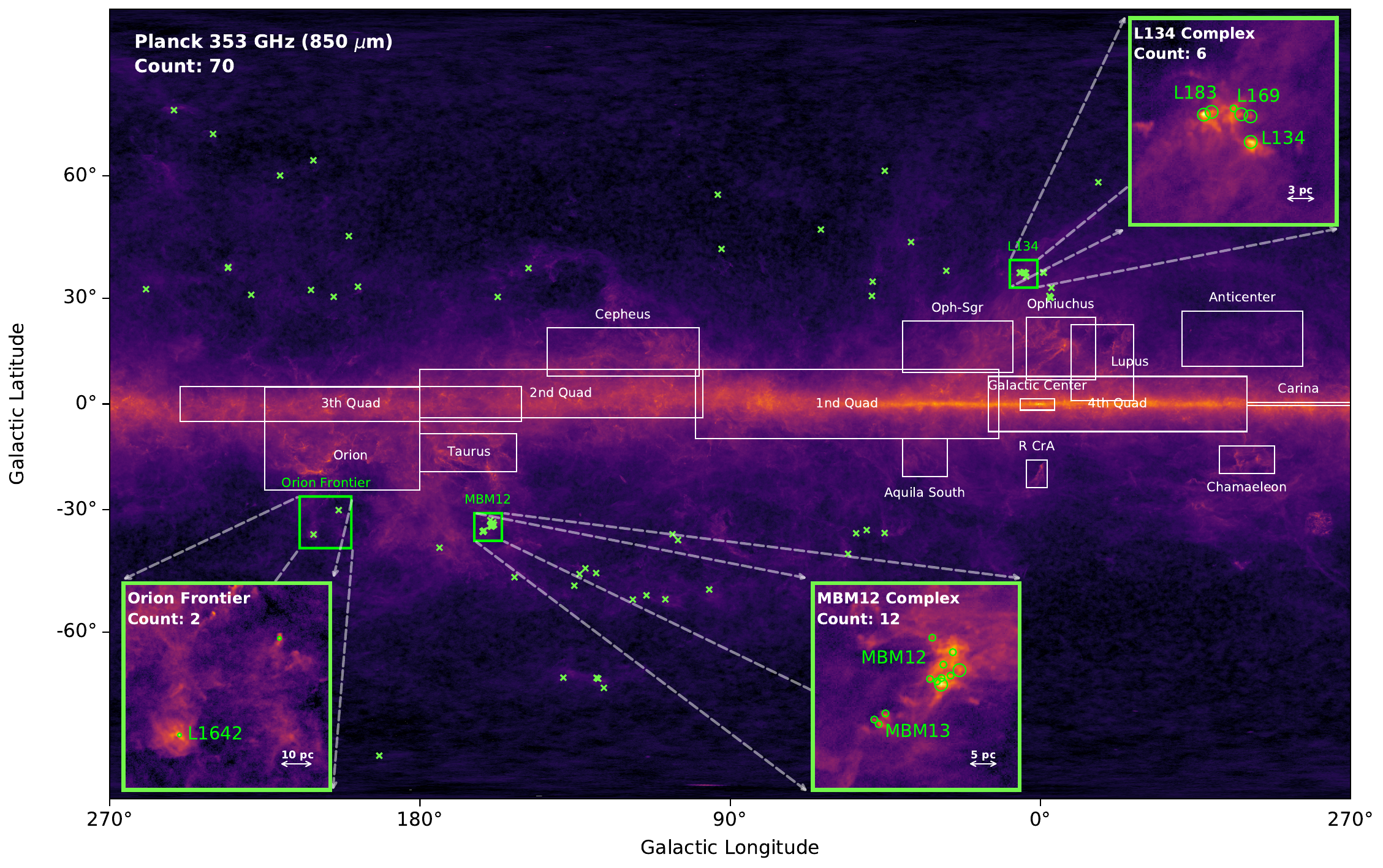}
\caption{The sky distribution of 70 high-latitude Planck Galactic cold clumps (HLPCs). The background color map is \textit{Planck} 353\,GHz (850\,$\mu$m) emission in cylinder projection. The 70 HLPCs are marked with green crosses. Three foreground subregions are zoomed in toward the Orion Frontier, the MBM 12 Complex, and the L134 Complex. The HLPCs in these regions cover the high column density end ($N_{\mathrm{H}_2}>1.0\times10^{21}$\,cm$^{-2}$). In the zoomed-in subregions, HLPCs are marked with green open circles whose sizes are equivalent to the size of the observing fields. The identifiers in each subregion are marked as green text. The names of the subregions and the number of HLPCs are marked with white text on the upper left. The white rectangles are the CO emission regions. \label{fig:HLP_Distribution}}
\end{figure*}

A thorough search of the SCUBA-2 850\,$\mu$m data in the JCMT Science Archive\footnote{\href{https://www.cadc-ccda.hia-iha.nrc-cnrc.gc.ca/en/jcmt/}{https://www.cadc-ccda.hia-iha.nrc-cnrc.gc.ca/en/jcmt/}}, and crossmatching with 793 HLPCs that satisfy the latitude criterion of $|b|>30\arcdeg$, gave 138 observing fields in total. We dismiss six of the fields that have limited integration time or nonstandard scan modes. Different scan patterns include constant velocity daisy patterns (CV Daisy) and rotating curvy pong patterns (Curvy Pong), so the field offsets vary significantly between different patterns. We make sure that the observing fields cover the peak of PGCCs at 353\,GHz. We also check superposition or repetition: if two fields cover the same PGCC, we choose the one with the higher sensitivity. After the work flow, a total of 70 SCUBA-2 fields are selected as the sample in this work. 

\subsection{Sample Properties} \label{data:property}

\begin{figure}
\centering
\includegraphics[width=0.98\linewidth]{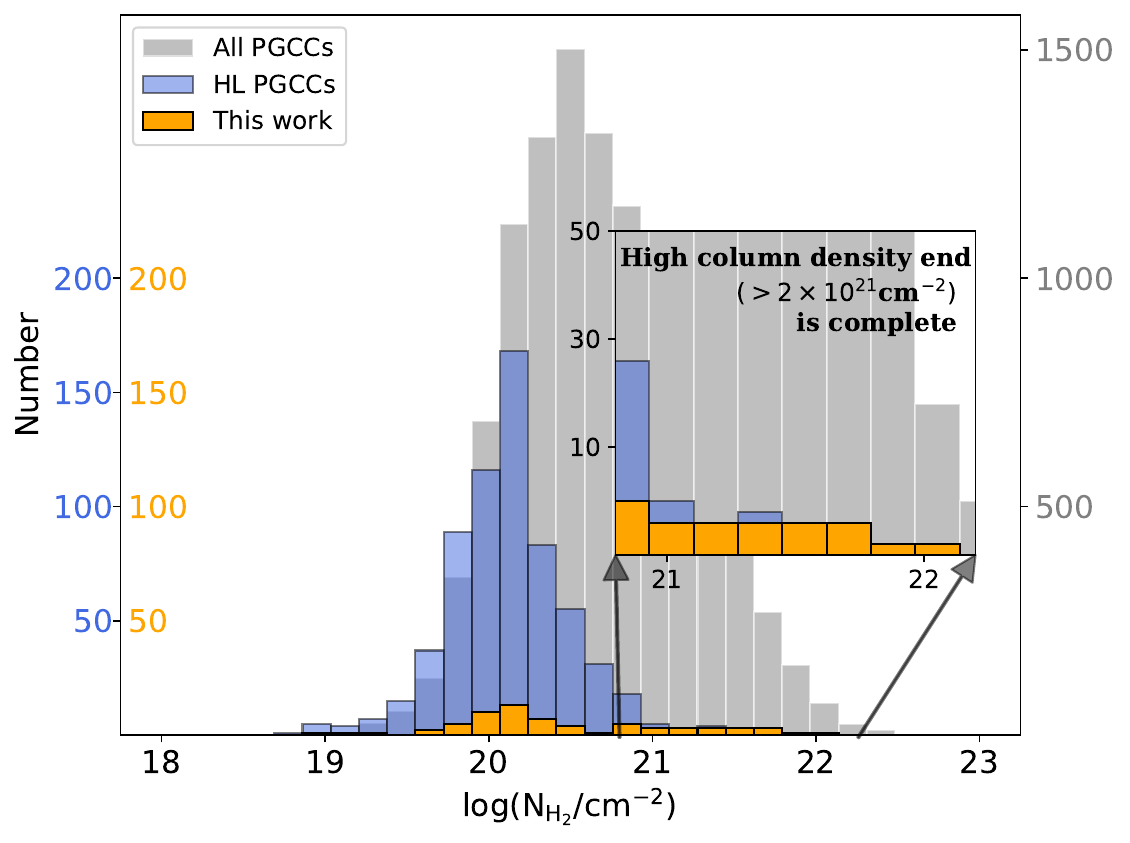}
\caption{The gray histogram shows the distribution of 13,188 PGCCs. The blue histogram is for 793 high-latitude PGCCs while the orange histogram is for the 70 HLPCs used in this work. A subpanel zooms in the H$_2$ column density range of $10^{20.8}$--$10^{22.2}$\,cm$^{-2}$. \label{fig:stats}}
\end{figure}

Seventy HLPCs are shown with green crosses, overlaid on the background \textit{Planck} 353\,GHz (850\,$\mu$m) continuum emission in Figure\,\ref{fig:HLP_Distribution}. White rectangles outline the CO emission regions defined by \citet{1987ApJ...322..706D}. The clump-averaged $N_{\rm H_2}$ H$_2$ column density $N_{\rm H_2}$ was calculated assuming a dust-emissivity model at 857\,GHz \citep{Planck2016XXVIII-PGCCs}. The $N_{\rm H_2}$ distributions of three samples--all the PGCCs, 793 high-latitude PGCCs, and 70 HLPCs--are plotted as gray, blue and orange histograms in Figure\,\ref{fig:stats}, respectively. 70 HLPCs has been evenly sampled in the $N_{\rm H_2}$ space, ensuring a similar distribution with its parent sample, 793 HL PGCCs. More importantly, the studied sample includes the densest clumps at the high-column-density end ($N_{\mathrm{H}_2}>2.0\times10^{21}$\,cm$^{-2}$). Considering that the denser clumps should be more likely to produce dense cores, we have covered the complete HLPCs where dense cores could form. 

Three regions with relatively higher column density, namely the Orion Frontier, the MBM 12 Complex, and the L134 Complex, are further zoomed in with subpanels in Figure\,\ref{fig:HLP_Distribution}. The HLPCs therein correspond to those at the high-density end as mentioned above. The Orion Frontier contains the dark cloud L1642, which together with the MBM 12 Complex are two famous HL clouds that have confirmed star-forming activity \citep{2014A&A...563A.125M}. The L134 Complex is another region containing several HLPCs, including the widely studied dark cloud L183 \citep{Lee1999Catalog,Lee2001Infall,Juvela2002L183,Pagani2003L183}. 

\subsection{Distance Estimation}
\label{data:distance}

Distance is always a difficult quantity to estimate in astronomy. Previous studies have estimated the distances of HL molecular clouds to be 100\,pc from the velocity dispersion and the scale height of an ensemble of clouds \citep{1984ApJ...282L...9B,1985ApJ...295..402M}. The star counting confirms that the HL molecular clouds are indeed nearby objects with upper limit ranges from 125 to 275\,pc \citep{1986A&A...168..271M}. Using Strömgren photometry, \citet{Franco1989Photometry} derive the distances of several HGaL clouds to be 100--230\,pc. Both the small $V_\mathrm{lsr}$ and the lack of a double-sine wave signature in the distribution on the $l-V_{lsr}$ (Galactic longitude $l$) plane demonstrate that HL molecular gas belongs to the local interstellar medium (ISM) and is too close to the Sun for Galactic rotation to modulate the velocities \citep{1996ApJS..106..447M}. 

Recently, the \textit{Gaia} satellite has provided new photometric measurements toward galactic stars \citep{2016A&A...595A...1G}. Together with 2MASS and Pan-STARRS 1 optical and near-infrared photometry, \textit{Gaia} DR2 parallaxes can help to infer distances and reddenings of $\sim$ 800 million stars. These stars trace the reddening on a small patch of the sky, along different lines of sight and different distance intervals, allowing us to build a 3D dust-reddening map \citep{2014ApJ...783..114G,2019ApJ...887...93G}. In a given direction, a jump of dust reddening is expected at a distance where there is a dust clump. Distances are estimated by this method and are listed in column (9) of Table\,\ref{tab:obs}, with an average value of $200\pm60$\,pc, indicating that HLPCs are mostly local ISM. Adopting Eq.\,1 in \citet{Xu2021HGaL} and considering that the Sun is 10\,pc above the Galactic midplane \citep{Griv2021Sun}, the altitude $z$ from the midplane (in units of parsecs) is calculated from $z=d\sin(b)+10$ where $d$ is the distance and $b$ is the latitude, and listed in column (10). We note that some fields may contain extragalactic objects, so the distance should be only for foreground Galactic dust. 

%%%%%%%%%%%%%%%%%%%%%%%%%%%%%%%%%%%%%%%%%%
%%%%%%%%%%%%%%%%% Result %%%%%%%%%%%%%%%%%
%%%%%%%%%%%%%%%%%%%%%%%%%%%%%%%%%%%%%%%%%%

\section{Results} \label{sec:results}

\subsection{Source Extraction} \label{results:extraction}

We adopt the \textit{dendrogram} algorithm \citep{Rosolowsky2008Dendrograms} to extract dense structures and then measure their integrated flux, peak flux, size, and position by 2D Gaussian fitting. The details of the algorithm parameter settings and the source extraction procedure are introduced in Appendix\,\ref{app:extract}. Within the 70 input fields, we have initially detected a total of 20 sources that belong to 15 SCUBA-2 observing fields. The field names and extracted sources are listed in columns (1)--(2) of Table\,\ref{tab:det}. Central coordinates, standard deviation of the deconvolved major and minor axes, integrated flux, and peak intensity are listed in columns (3)--(7). 

\subsection{Source Identification and Dense Core Definition} \label{results:identification}

The CV Daisy mode of observation can produce artifacts \citep{Liu2018TOP-SCOPE,Eden2019SCOPE} that are extracted by the algorithm as false source detections. Therefore, we additionally require that both the peak intensity after being smoothed to the \textit{Planck} beam and the total flux of the source at 353\,GHz, are lower than those of the parent PGCC. This results in two sources, G50.41-35.40 SMM1 and G197.98+33.10 SMM1, being classified as artifacts and excluded from further analyses. We note that the flux given by the PGCC catalog should be from the cold residual map \citep{Planck2016XXVIII-PGCCs}, so the original \textit{Planck} flux at 353 GHz could be even larger. However, considering the cold nature of dense cores, this should contribute little to the warm component.

We crossmatch the true detections within 1\arcmin~using SIMBAD\footnote{\href{http://simbad.u-strasbg.fr/simbad/}{http://simbad.u-strasbg.fr/simbad/}}. We find that only one field, G6.04+36.77, pointing toward the molecular cloud L183, contains three resolved (or marginally resolved) sources, which were previously identified as low-mass prestellar cores \citep{Dickens2000L183,Pagani2003L183}. 
Our other detections are unresolved as point sources, classified as either young stellar object (YSO) or extragalactic object (point sources) -- including gravitational lensed galaxy (LeG), BL Lacertae object (BLL), protocluster of galaxies (PClG), active nuclei candidate (AGN), and quasar (QSO). The determined identifiers and references are given in columns (8) and (9) of Table\,\ref{tab:det}.

The physical parameters of the L183 dense cores are calculated in Appendix\,\ref{app:psc}. Throughout the paper, we adopt the empirical definition of dense core with typical size of $\lesssim0.1$\,pc and density of $10^{5}$--$10^{6}$\,cm$^{-3}$ \citep{WT1994JCMT,WT1999IRAM30m,Kirk2005JCMT}.

%%%%%%%%%%%%%%%%%%%%%%%%%%%%%%%%%%%%%%%%%%
%%%%%%%%%%%%%%% Discussion %%%%%%%%%%%%%%%
%%%%%%%%%%%%%%%%%%%%%%%%%%%%%%%%%%%%%%%%%%

\section{Discussion} \label{sec:discuss}

\subsection{Dense Cores Are Scarce In HLPCs} \label{discuss:scarcity}

Having only one detection among the 70 HLPCs highlights the scarcity of dense cores at these latitudes. To further confirm this discrepancy compared to the low-latitude ($|b|<30\arcdeg$) counterpart, 1235 observing fields from the JCMT Large Project ``SCUBA-2 Continuum Observations of Pre-protostellar Evolution'' \citep[SCOPE;][]{Liu2018TOP-SCOPE,Eden2019SCOPE} are used as a comparison group because of the following two reasons: (1) twenty-one HLPC fields in this work come from the SCOPE project so that they have been observed with comparable sensitivities; (2) the SCOPE observations serve as a representative sample of PGCCs, with similar distributions in distance, size, and temperature, and with complete column density coverage over $10^{21}$ cm$^{-2}$ \citep{Liu2018TOP-SCOPE}. 

Considering the beam dilution effects for marginally resolved sources, the column densities are corrected by a beam filling factor $B_{\rm ff}=(\theta_{i}^2)/\theta_{\rm PSF}^2$ where $\theta_{\rm PSF}=4\parcmin3$ is the \textit{Planck} beam size at 857\,GHz \citep{Planck2016XXVIII-PGCCs} and $\theta_{i}$ is the intrinsic size deconvolved from the beam. For extended sources of which intrinsic size exceeds the beam size, $B_{\rm ff}$ is set to 1, and no correction is performed. As a result, the corrected column density $N^{\prime}_{\rm H_2}$ increases by a factor of 1.1 on average and 9 at maximum. 

\begin{figure*}[!t]
\centering
\includegraphics[width=0.8\linewidth]{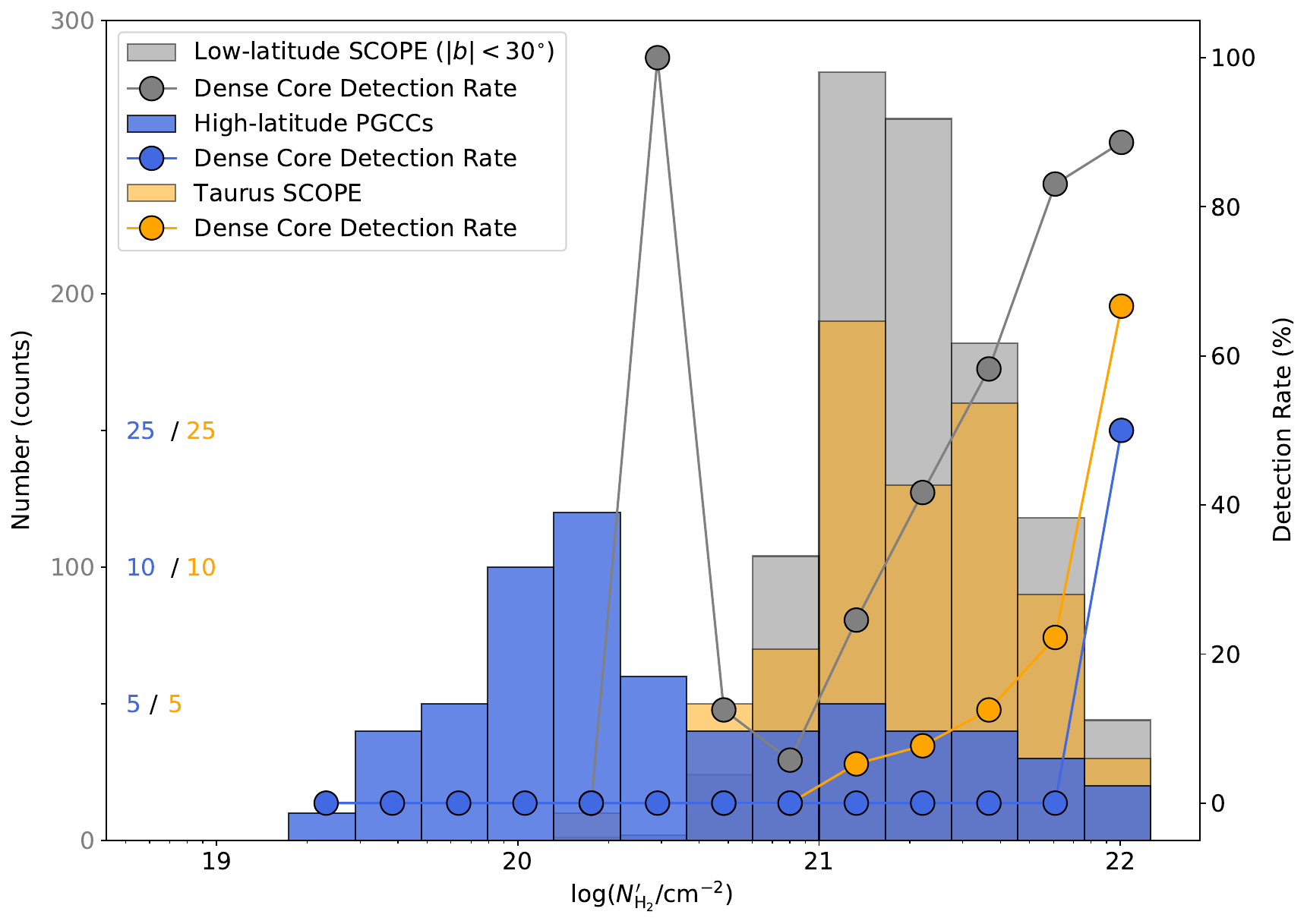}
\caption{The distribution of column density corrected by beam filling factor $N^{\prime}_{\rm H_2}$} for low-latitude SCOPE fields (gray), the HLPC fields (blue) and Taurus SCOPE fields (orange), respectively. The connected data points in corresponding colors depict the DCDR in various $N_{\rm H_2}$ bins. \label{fig:dr_stats}
\end{figure*}

In Figure\,\ref{fig:dr_stats}, we present the number distributions of different samples across a set of $N^{\prime}_{\rm H_2}$ bins, denoted as ${S_{{\rm samp},i}}$ where $i$ indicates bin index. Specifically, the distributions for the low-latitude SCOPE fields and the HLPC fields are depicted using the gray and blue histograms, respectively. We also collect the number distribution of those fields with dense cores detected $\{S_{{\rm det},i}\}$. The dense core detection rate (DCDR), defined as $\{S_{{\rm det},i}/S_{{\rm samp},i}\}$, is shown with connected data points. For the low-latitude SCOPE, the DCDR experiences a pronounced increase at a threshold of column density around $N_{\rm H_2}\simeq1.0\times10^{21}$\,cm$^{-2}$, reaching 90\% at the $N_{\rm H_2}\simeq5.0\times10^{21}$\,cm$^{-2}$ regime. The column density threshold for forming dense cores is consistent with what has been found in Gould Belt clouds \citep[e.g.,][]{Johnstone2004Threshold}. The sudden peak at $3\times10^{20}$\,cm$^{-2}$ likely results from the limited sample size. In contrast to the DCDR of the low-latitude SCOPE clumps, the DCDR for the HLPCs remains zero for $N_{\rm H_2} < 1.0\times10^{22}$\,cm$^{-2}$, until dense cores in the HLPC G6.04+36.77 are detected. 

The SCOPE serve as a gauge to tell whether the HLPCs have significantly scarce dense cores. Given the null hypothesis that the HLPCs share a DCDR greater or equal to that of the SCOPE, the number of the HLPC fields containing dense cores in each bin of $N_{\rm H_2}\geq 10^{21}$\,cm$^{-2}$ can be predicted, as $\{S_{{\rm pred},i}\}_{\rm HLPC}$. The one-sided Mann–Whitney U test \citep{Mann-Whiteney-U-test} is performed between the $\{S_{{\rm pred},i}\}_{\rm HLPC}$ and $\{S_{{\rm det},i}\}_{\rm HLPC}$, giving a p-value of $4.8\times10^{-3}$. This is $\ll0.05$, thus robustly excluding the null hypothesis. In other words, dense cores are scarce in HLPCs compared to the low latitude.

% We also check whether a difference in distance causes bias. The average distance for the Taurus clumps is approximately 180\,pc , which closely aligns with the corresponding value for the HLPCs. The Taurus clumps covered by SCOPE are extracted as a subsample, Taurus SCOPE. Figure\,3 presents the number distribution and the DCDR of the Taurus SCOPE in orange. In the same way as above, the Mann-Whitney U test gives a p-value of $0.042<0.05$, thus favoring that the DCDR of the Taurus clumps is statistically distinguishable from that of HLPCs. As a result, by conducting a comparison between the HLPCs and the Taurus SCOPE clumps, which share an equivalent mean distance, the caveat from distance is excluded.

% We notice that although Taurus serves as a calibrator for distance, the DCDR of Taurus is systematically lower than that of other low-latitude sources, although the Mann-Whitney U test gives a moderate p-value of 0.07. See further discussion in Section 4.3. 

\subsection{What Does ``Dense'' Mean?} \label{discuss:dense}

The scarcity of detection does not necessarily indicate the scarcity of dense cores, primarily due to two factors: (1) the limited sensitivity of identifying sources and (2) the absence of large-scale flux in the SCUBA-2 data processing. Therefore, to understand what the scarcity of dense cores means, it is essential to clarify what ``dense'' means. 

\begin{figure*}[!t]
\centering
\includegraphics[width=0.8\linewidth]{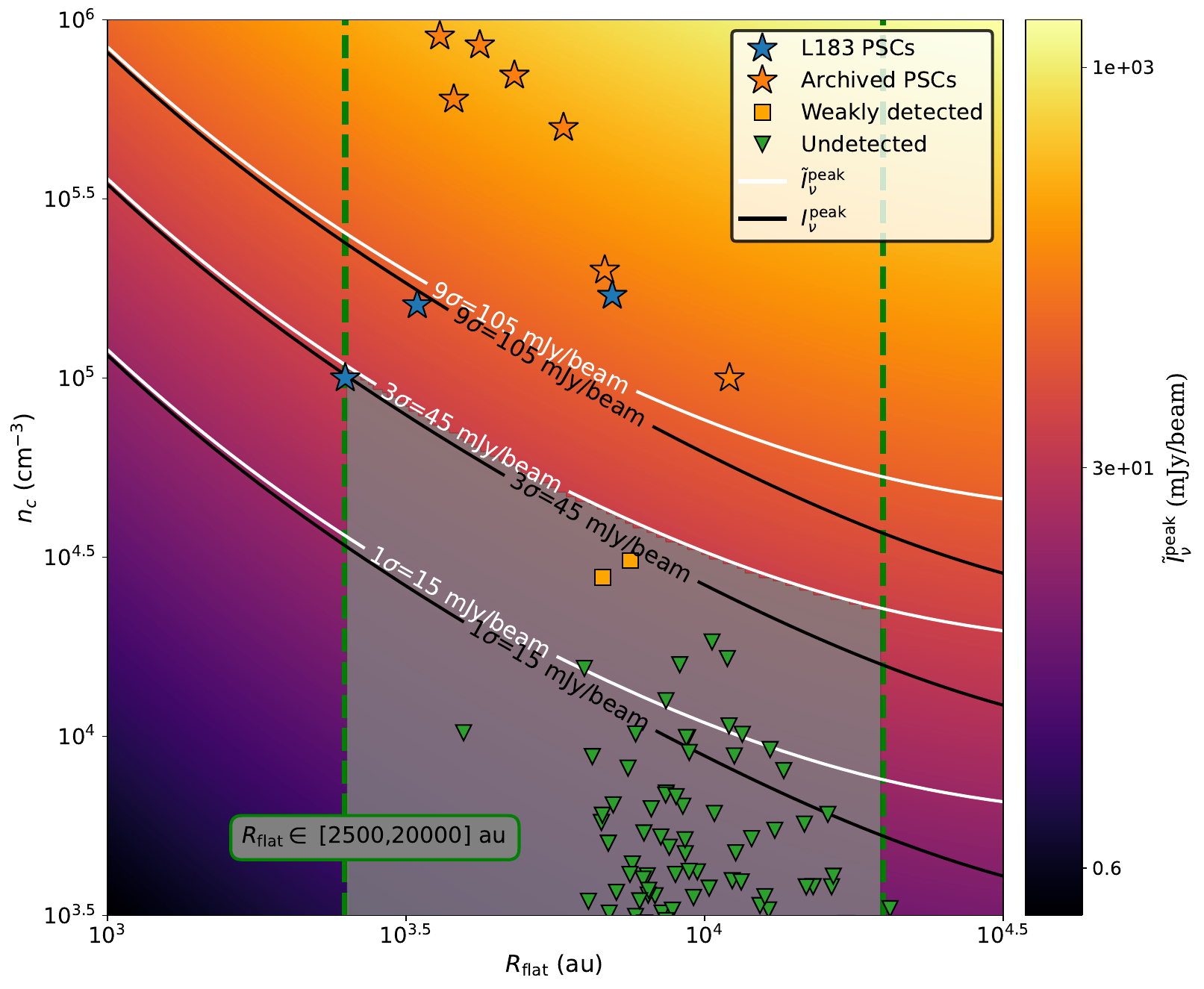}
\caption{The peak intensity $\widetilde{I}^{\rm peak}_{\nu}$ observed by SCUBA-2 across parameter space defined by the flat radius $R_{\rm flat}\in[10^{3},10^{4.5}]$\,au and the central density $n_c\in[10^{3.5},10^{6}]$\,cm$^{-3}$. The white curves mark $1\sigma$, $3\sigma$, and $9\sigma$ levels, while the black curves trace the same loci of synthetic model (no large-scale flux filtered out). The green dashed lines delineate the flat inner radius limits $R_{\rm flat}\in[2500,20000]$. The shaded gray region indicates the permissible parameter interval of undetected prestellar cores should they exist. Blue and red stars show the L183 prestellar cores and other previously detected prestellar cores \citep{WT1994JCMT,WT1999IRAM30m,Kirk2005JCMT}. The orange rectangles show two weakly detected cores in our fields. The green triangles show cores detected by Herschel in several HLPCs \citep{Montillaud2015GCCs}, but undetected by SCUBA-2. \label{fig:syn}}
\end{figure*} 

Prestellar cores are observed to have flat inner-density gradients that approach $\rho\sim r^{-2}$ beyond a few thousand astronomical units \citep{WT1994JCMT,WT1999IRAM30m,Kirk2005JCMT}, which can be reproduced by a nonmagnetic and Plummer-like model \citep{Whitworth2001L1544} as
\begin{equation} \label{eq:density}
n_{\rm H_2}(r) = \frac{n_{c}}{1+(r/R_{\rm flat})^2},
\end{equation}
where $n_{c}$ is the central H$_2$ number density and $R_{\rm flat}$ is the flat inner radius. The column density profile of such a model core has the analytical form of
\begin{equation} \label{eq:column}
N_{\rm H_2}(p) = \frac{2n_c R_{\rm flat}}{(1+p^2/R_{\rm flat}^2)^{1/2}}\times \tan^{-1}\left[\frac{(R_{\rm out}^2-p^2)^{1/2}}{(R_{\rm flat}^2+p^2)^{1/2}}\right], 
\end{equation}
where $p$ is the distance from core center in the plane of the sky \citep{Dapp2009PSC} and $R_{\rm out}=0.2$\,pc $\simeq40000$\,au defines the boundary of core. 

Using the combined ammonia data from the Green Bank Telescope and Karl G. Jansky Very Large Array, the temperatures of the three prestellar cores are observed to have a minor decrease toward the center of the core $\lesssim2000$\,au \citep{Lin2023PSC}. Therefore, we consider a constant temperature profile as $T(r)=T_0=10$\,K in the following discussion. 

Assuming optically thin dust emission, the column density can be used to synthesize model intensity as, 
\begin{equation} \label{eq:I850}
I_{\nu}(p) = \frac{N_{\rm H_2}(p) \Omega\mu_{\rm H_2}m_{\rm H} \kappa_{\nu}B_{\nu}(T_{\rm dust})}{\mathcal{R}},  
\end{equation}
where $\Omega = \pi\theta_{\rm beam}^2/4\ln2$ measures the solid angle (in unit of radian) per JCMT beam (with FWHM of $\theta_{\rm beam}$), $\mu_{\rm H_2} = 2.81$ is the molecular weight per hydrogen molecule \citep{Evans2022SlowSF}, $m_{\rm H}$ is the mass of a hydrogen atom, $\kappa_{\nu}=1.22$\,cm$^{2}$\,g$^{-1}$ \citep{Beckwith1990YSO} is the dust opacity at frequency of $\nu=350$\,GHz ($\sim850$\,$\mu$m), $B_{\nu} (T_{\rm dust})$ is the Planck function at a given dust temperature $T_{\rm dust}$, and $\mathcal{R}=100$ is the gas-to-dust mass ratio. 

Now we synthesize the SCUBA-2 image $\widetilde{I}_\nu(x,y)$ by adding a high-frequency filter $\sqrt{u^2+v^2}>\xi$, where $\xi$ corresponds to $200\arcsec$ in the frequency space \citep{Mairs2015GBS},
\begin{equation} \label{eq:filter}
\begin{aligned}
& \widetilde{I}_\nu(x,y) = \iint_{\{\sqrt{u^2+v^2}>\xi\}} J_\nu (u,v) e^{i2\pi(ux+vy)} \mathrm{d}u\,\mathrm{d}v,
\end{aligned}
\end{equation}
where $J_\nu (u,v)=\iint I_\nu (x,y) e^{-i2\pi(ux+vy)} \mathrm{d}x\mathrm{d}y$ is the synthetic model in the Fourier frequency space. As a result, the synthetic observed intensity $\widetilde{I}_\nu(x,y)$ can simulate the observed large-scale missing flux at the SCUBA-2 data processing. 

In Figure\,\ref{fig:syn}, the background color map shows observed peak intensities $\widetilde{I}^{\rm peak}_{\nu}=\widetilde{I}_{\nu}(0,0)$ across the 2D parameter space of flat radius $R_{\rm flat}\in[10^{3},10^{4.5}]$\,au and central density {$n_c\in[10^{3.5},10^{6}]$\,cm$^{-3}$. The prestellar cores have been reported to have $R_{\rm flat}>2500$\,au \citep{WT1999IRAM30m,Kirk2005JCMT}, which are delineated by the left green dashed line in Figure\,\ref{fig:syn}. The right green dashed line marks 20,000\,au, which corresponds to 0.1\,pc. 

Consistent with the criteria in the source extraction algorithm (see Appendix\,\ref{app:extract}), a threshold of $3\sigma$ is adopted to constrain the upper limit of $\widetilde{I}^{\rm peak}_{\nu}$. As a result, the gray shaded region traces the permissible parameter interval for an undetected core in our observations. In other words, if such cores exist, they should have $n_c<10^{5}$\,cm$^{-3}$, which is considerably less dense than those that have been identified in nearby low-mass cloud cores \citep{WT1999IRAM30m,Kirk2005JCMT}. 

To further demonstrate this upper limit of density, we smooth the images to a resolution of 20\arcsec. With better sensitivity, two new cores (on in HLPC G159.41-34.37 and one in G161.87-35.76) are identified by the same algorithm and parameter input, which are called weakly detected cores. They have radii of about 6700 and 7500 au and averaged density of $2.8\times10^4$ cm$^{-3}$ and $3.1\times10^4$ cm$^{-3}$, respectively. The two cores are labeled as orange rectangles in Figure\,\ref{fig:syn}. We also retrieve \textit{Herschel} cold cores in L134 (HLPCs G4.13+35.75 and G4.17+36.67 in our survey), MBM12 (G159.21-34.28, G158.51-33.99, G159.14-33.79, G159.23-34.51, G159.41-34.37, G159.66-34.31), L1642 (G210.90-36.55), and LDN1780 (G358.96+36.81, G359.21+36.89) from \citet{Montillaud2015GCCs} which are labeled as green triangles. As they all lie in the gray region which is below the sensitivity limit, these cold cores are not dense enough for detection and are consistently below the density limit of $10^{5}$\,cm$^{-3}$. As noted by \citet{WT2016L1495}, SCUBA-2 selects the densest cores from a population at a given temperature, which makes SCUBA-2 ideal for identifying those cores in Herschel catalogs that are closest to forming stars. So it is of great interest to study how these low-density cores form and whether they can still form stars or are transient objects. 

As indicated by the black curves, peak intensity of a synthetic model $I^{\rm peak}_\nu$ is always below the corresponding $\widetilde{I}^{\rm peak}_{\nu}$ outlined by white curves. The difference reflects the missing large-scale flux, which increases in importance from 0.06 to 0.31 dex with $R_{\rm flat}$ from 2500 to 20,000\,au. Therefore, if we do not consider missing flux, the density limit can be even lower, especially for those cores with larger flat radius.

\subsection{Star Formation at High Galactic Latitude} \label{discuss:physics}

Low density and high virial parameter lead to a challenge for direct gravitational collapse and then star formation of HL clouds. Observationally, it is consistent with infrared cirrus which is thought to be hostile to star formation \citep{Low1984Cirrus} and the dispersed populations of pre-main-sequence stars \citep[see review by][]{McGehee_HGaL_Review}. Recently, a clear decreasing trend of N$_2$H$^+$\,(1–0) and C$_2$H\,(1–0) detection rates with latitude is found by \citet{Xu2021HGaL}. Besides, HCN\,(1–0) and HCO$^+$\,(1–0) line survey by \citet{Braine2023Outer} reveals that HL molecular clouds have lower dense gas fractions compared to those in the Galactic plane. Theoretically, based on Jeans mass arguments, these low-density turbulent clouds have molecular gas mass lower than the turbulent Jeans mass \citep[see Table 5 in][]{Xu2021HGaL}, therefore unable to fragment into dense dust cores, or protostellar embryos, which agrees with the scarcity of dense cores observed by SCUBA-2. 

Previous studies have reported that the virial parameters of the PGCCs in the Taurus region (Taurus clumps hereafter) are predominantly greater than 1, with a median value of approximately 9 \citep{Meng2013Taurus}. This value is considerably lower than the median virial parameter of HLPCs, which stands at about 35. The Taurus clumps covered by the SCOPE project are designated as the Taurus SCOPE subsample. Figure \ref{fig:dr_stats} displays the number distribution and the DCDR for the Taurus SCOPE in orange. In the same way as above, the Mann-Whitney U test gives a $p$-value of $0.042<0.05$, thus favoring that the DCDR of the Taurus clumps is statistically larger than that of HLPCs. Interestingly, the Taurus clumps also exhibit a significantly smaller DCDR compared to the low-latitude SCOPE clumps, as evidenced by a Mann-Whitney U test p-value of $5.4\times10^{-3}$. This indicates that dense cores within the Taurus clumps are relatively rarer compared to other SCOPE clumps. Consequently, the Taurus clumps occupy an intermediate position between the HLPCs and low-latitude SCOPE clumps in terms of DCDR and virial parameter. The observed trend of decreasing DCDR with increasing virial parameter further substantiates the link between core formation efficiency and the dynamic state of the gas, as previously suggested \citep{Eden2019SCOPE}. 

HLPCs have a distance of 200 pc which is highly consistent with the radius of the Local Bubble (LB) created by supernovae \citep{Zucker2022LocalBubble}. The LB is reported to expand and sweep up the ambient interstellar medium into a shell that has now fragmented and collapsed into the most prominent nearby molecular clouds.  Interestingly, \citet{Zucker2022LocalBubble} also found that the Taurus star formation region is very likely being compressed by two super bubbles: the local super bubble and the smaller Per-Tau super bubble. If so, it is probable that the formation of dense cores can be hindered by supernova shocks in the solar neighborhood. If the HLPCs and the Taurus clumps were on the shell of LB, the scarcity of dense cores should favor turbulence-inhibited rather than supernova-driven star formation. 

On the other hand, the scarcity itself is what gives the only detection (L183) in our survey, as well as a few other clouds (such as MBM 12 and 20), unique status. The capacity of these high-latitude clouds to form cold molecular cores and young stars could arise from a confluence of conditions including variations in the interstellar radiation field, changes in dust grain size and chemistry, the occurrence of shocks, and transient events in the ISM \citep{McGehee_HGaL_Review}. Consequently, in-depth explorations of the physical and chemical processes within these high-latitude dense cores, for example the L183 dense cores, are merited. 

\section{Conclusion} \label{sec:conclude}

We performed a JCMT SCUBA-2 archival investigation of 70 fields toward HLPCs to search for dense cores. The sample benefits from being representative of the total HLPC population at low column density ($<2\times10^{21}$\,cm$^{-2}$) and covering the densest clumps at the high column density end ($>1\times10^{21}$\,cm$^{-2}$). Using dust reddening in 3D map, the distances of the HLPCs are estimated to be 110--410\,pc with a mean value of 200($\pm$60) pc. A total of 17 SCUBA-2 sources are identified from a mean noise rms of 15\,\mjybeam. Only one field G6.04+36.77 (L183) contains three dense Galactic cores. The other 14 unresolved sources include 12 extragalactic objects and two Galactic YSOs.

Compared to the low-latitude SCOPE clumps and the Taurus clumps (at a similar distance to HLPCs), the DCDR of HLPCs is significantly lower at the high column density end ($>1\times10^{21}$\,cm$^{-2}$). 
Statistical tests verify the scarcity of dense cores in HLPCs. With synthetic observations of known dense cores, the central density of any undetected dense cores is constrained to be $n_c\lesssim10^5$\,cm$^{-3}$, should they exist in HLPCs. The observed scarcity of dense cores aligns with the low-density turbulent environment in HLPCs, as proposed in previous far-infrared and CO line surveys. If the HLPCs and the Taurus clumps were on the shell of the Local Bubble, the scarcity of dense cores should favor turbulence-inhibited rather than supernova-driven star formation. Furthermore, the scarcity also calls for further study on the formation mechanism of L183 dense cores. 

\section*{Acknowledgment}

This work has been supported by the National Key R\&D Program of China (No. 2022YFA1603102, 2019YFA0405100), the National Science Foundation of China (12033005, 11973013), and the China Manned Space Project (CMS-CSST-2021-A09, CMS-CSST-2021-B06), and the China-Chile Joint Research Fund (CCJRF No. 2211). CCJRF is provided by Chinese Academy of Sciences South America Center for Astronomy (CASSACA) and established by National Astronomical Observatories, Chinese Academy of Sciences (NAOC) and Chilean Astronomy Society (SOCHIAS) to support China-Chile collaborations in astronomy. 
T.L. acknowledges the support by the international partnership program of Chinese academy of sciences through grant No.114231KYSB20200009, and Shanghai Pujiang Program 20PJ1415500. 
This research was carried out in part at the Jet Propulsion Laboratory, which is operated by the California Institute of Technology under a contract with the National Aeronautics and Space Administration (80NM0018D0004). 
This work is sponsored (in part) by the Chinese Academy of Sciences (CAS), through a grant to the CAS South America Center for Astronomy (CASSACA) in Santiago, Chile. 
D.J. is supported by NRC Canada and by an NSERC Discovery grant. 
G.G. acknowledges support from the ANID BASAL project FB210003.
E.F. acknowledges support from the European Council, under the European Community's Seventh framework Programme, through the Advance Grant MIST (FP7/2017-2024, No. 742719).
P.S. was partially supported by a Grant-in-Aid for Scientific Research (KAKENHI Number JP22H01271 and JP23H01221) of JSPS. 
C.W.L. is supported by the Basic Science Research Program through the National Research Foundation of Korea (NRF) funded by the Ministry of Education, Science and Technology (NRF-2019R1A2C1010851), and by the Korea Astronomy and Space Science Institute grant funded by the Korea government (MSIT) (Project No. 2023-1-84000). 
G.J.W. gratefully acknowledges receipt of an Emeritus Fellowship from The Leverhulme Trust. 
M.J. acknowledges support from the Research Council of Finland grant 348342.
The work of M.G.R. is supported by NOIRLab, which is managed by the Association of Universities for Research in Astronomy (AURA) under a cooperative agreement with the National Science Foundation. 
N.H. acknowledges support from the Nation Science and Technology Council (NSTC) of Taiwan with grant NSTC 111-2112-M-001-060.SPL acknowledges the Ministry of Science and Technology of Taiwan for grant 112-2112-M-007 -011.
% JCMT acknowledgement part 
This research used the facilities of the Canadian Astronomy Data Centre operated by the National Research Council of Canada with the support of the Canadian Space Agency. The James Clerk Maxwell Telescope is operated by the East Asian Observatory on behalf of The National Astronomical Observatory of Japan; Academia Sinica Institute of Astronomy and Astrophysics; the Korea Astronomy and Space Science Institute; the National Astronomical Research Institute of Thailand; Center for Astronomical Mega-Science (as well as the National Key R\&D Program of China with No. 2017YFA0402700). Additional funding support is provided by the Science and Technology Facilities Council of the United Kingdom and participating universities and organizations in the United Kingdom and Canada. Additional funds for the construction of SCUBA-2 were provided by the Canada Foundation for Innovation.

% Software acknowledgement part 
\software{Astropy, a community developed core python package for astronomy \citep{Astropy2013,Astropy2018,Astropy2022}. Montage, funded by the National Science Foundation under Grant Number ACI-1440620, and previously funded by the National Aeronautics and Space Administration's Earth Science Technology Office, Computation Technologies Project, under Cooperative Agreement Number NCC5-626 between NASA and the California Institute of Technology \citep{Jacob2010Montage,Berriman2017Montage}.}

% Table 1: for the detected source catalog
\begin{deluxetable*}{ccccccccc}
% \rotate % rotate the wide table
% \centerwidetable
% \tabletypesize{\scriptsize}
% \tabletypesize{\small}
\tablewidth{0pt}
\linespread{1.3}
%\tablenum{1}
\tablecaption{Detected Sources Catalog \label{tab:det}}
\tablehead{
\colhead{Field} & \colhead{Source} & \colhead{RA} & \colhead{DEC} & \colhead{$\sigma_\mathrm{maj}\times\sigma_\mathrm{min}$} & \colhead{$F_\mathrm{int}$} & \colhead{$I_\mathrm{peak}$} & \colhead{Identifier$^{[\mathrm{refs}]a}$} & \colhead{Class$^{b}$} \\
\colhead{} & \colhead{} & \colhead{deg} & \colhead{deg} & \colhead{arcsec$^2$} & \colhead{$\mathrm{Jy}$} & \mjybeam & \colhead{} & \colhead{}
}
\colnumbers
\startdata
G6.04+36.77 & SMM1 & 238.5361 & -2.8732 & 34.2$\times$17.2 & 3.36 & 218.4 & \addstackgap[0pt]{\Centerstack[c]{Position~C$^{[1]}$ Region~3$^{[2]}$}} & PSC \\
            & SMM2 & 238.5025 & -2.8786 & 10.2$\times$7.4 & 0.13 & 57.4 & \addstackgap[0pt]{\Centerstack[c]{Position~W$^{[1]}$ Region~4$^{[2]}$}} & PSC \\
            & SMM3 & 238.5404 & -2.8154 & 13.5$\times$10.0 & 0.35 & 64.2 & \addstackgap[0pt]{\Centerstack[c]{Position~N$^{[1]}$ Region~5$^{[2]}$}} & PSC \\
G45.12+61.11 & SMM1 & 225.6511 & 29.3460 & point & 0.08 & 111.2 & \addstackgap[0pt]{\Centerstack[c]{PLCK~G045.1+61.1$^{[3,4]}$}} & LeG \\
G50.41-35.40 & SMM1 & 321.7839 & -2.7151 & 10.3$\times$8.3& 0.52 & 148.5 & - & Artifact \\
G53.44-36.25 & SMM1 & 323.7980 & -1.0478 & point & 0.09 & 111.8 & SMMJ2135-0102$^{[5]}$ & LeG \\
G92.49+42.88 & SMM1 & 242.3232 & 60.7542 & point & 0.08 & 110.4 & PLCK~G092.5+42.9$^{[3,4]}$ & LeG \\
% G128.94-46.39 & SMM1 & 17.1083 & 16.3747 & UR & 0.19 & 272.9 & - & Artifact \\
G152.54-47.36 & SMM1 & 32.8050 & 10.8598 & point & 0.89 & 942.0 & J021113.1+105134$^{[6]}$ & BLL \\
G157.44+30.33 & SMM1 & 113.3787 & 117.2158 & point & 0.07 & 74.8 & PLCKESZ$^{[7]}$ & ClG \\
G197.98+33.10 & SMM1 & 128.3946 & 26.1982 & 36.8$\times$27.7 & 6.29 & 176.6 & - & Artifact \\
G200.62+46.09 & SMM1 & 143.0981 & 27.4163 & point & 0.06 & 75.1 & PLCK~G200.6+46.1$^{[3]}$ & LeG \\
G204.99+30.38 & SMM1 & 127.6930 & 19.6131 & point & 0.014 & 20.5 & Planck18p194-0$^{[8]}$ & PClG \\
              & SMM2 & 127.7268 & 19.6251 & point & 0.009 & 15.4 & Planck18p194-1$^{[8]}$ & PClG \\
              & SMM3 & 127.6705 & 19.6631 & point & 0.008 & 14.2 & Planck18p194-3$^{[8]}$ & PClG \\
G210.90-36.55 & SMM1 & 68.7600 & -14.2287 & point & 0.05 & 64.1 & \addstackgap[0pt]{\Centerstack[c]{GCVS~EW~Eri$^{[9]}$ MJR2015~1752$^{[10]}$}} & \addstackgap[0pt]{\Centerstack[c]{V* Y*O}} \\
              & SMM2 & 68.7080 & -14.2195 & point & 0.08 & 82.1 & \addstackgap[0pt]{\Centerstack[c]{HH123$^{[11]}$ MJR2015~1751$^{[10]}$}} & \addstackgap[0pt]{\Centerstack[c]{HH Y*O}} \\
G211.62+32.23 & SMM1 & 131.7098 & 15.0943 & point & 0.11 & 125.8 & J084650.1+150547$^{[12]}$ & AGN? \\
G228.99+30.91 & SMM1 & 137.2924 & 1.3599 & point & 0.23 & 274.9 & 4C~01.24B$^{[13]}$ & QSO \\
G343.12+58.61 & SMM1 & 212.5190 & 2.0516 & point & 0.04 & 68.4 & J141004.6+020306$^{[14]}$ & BLL \\
\enddata
\tablecomments{The HLPC fields are listed in column (1). The extracted sources are named as SMM$\mathcal{X}$, as listed in column (2). The equatorial coordinates of R.A. and Dec. in Epoch J2000 are listed in columns (3) and (4). The deconvolved standard deviation along the major and minor axis are listed in columns (5), and marked as ``point'' if the source is unresolved as a point source. The integrated flux and peak intensity are listed in columns (6) and (7). Identifier(s) and classifications retrieved from SIMBAD are listed in columns (8) and (9).}
\tablenotetext{a}{ 
[1] \citet{Dickens2000L183}; % define four positions (C,W,N,S) with local emission peaks at different molecules; 
[2] \citet{Karoly2020L183}; % define five regions (1--5) in Stokes I map by JCMT SCUBA-2/POL-2 at 850\,$\mu$m; 
[3] \citet{Canameras2015GEMS}; % present CO spectroscopy and infrared-to-millimetre dust photometry of 11 exceptionally bright gravitationally lensed galaxies; 
[4] \citet{Frye2019G165}; % present Hubble Space Telescope WFC3-IR imaging in the fields of six apparently bright dusty star-forming galaxies; 
[5] \citet{Swinbank2010Galaxy}; % use a gravitationally lense with 32 time magnifier to resolve star-forming regions in a z=2.3 galaxy; 
[6] \citet{Healey2008BLLac}; % provide a large catalog of likely Gamma-ray active galactic nuclei, i.e., CGRaBS; 
[7] \citet{Planck2014XXIX}; % provide Planck Sunyaev-Zeldovich sources catalog; 
[8] \citet{MacKenzie2017SCUBA2}; % a SCUBA-2 survey of 61 Planck High-Z, where ten are identified as lensed galaxies and 51 are called "Planck overdensities"
[9] \citet{Samus2017Vstar}; % provide a catalog of variable stars; 
[10] \citet{Montillaud2015GCCs}; % provide a catalog of Planck Galactic cold cores observed by Herschel; 
[11] \citet{Reipurth1999HH123}; % identify HH object
[12] \citet{Truebenbach2017AGN}; % an invisible Active Galactic Nucleus Candidate Catalogue: a mid-infrared-radio selection method for optically faint active galactic nuclei.
[13] \citet{Wright2009Catalog}; % the list of point sources found in the Wilkinson Microwave Anisotropy Probe (WMAP) five-year maps
[14] \citet{Plotkin2008BLLac}; % a large sample of BL Lac objects from the SDSS and FIRST.
}
\tablenotetext{b}{Classification according to references. PSC--prestellar core. LeG--gravitational lensed galaxy. Artifact--probable artifact by JCMT data reduction pipeline. BLL--BL Lacertae object. ClG--cluster of galaxies. PClG--proto-cluster of galaxies. V*--variable star. Y*O--young stellar object. HH--Herbig-Haro object. AGN?--active galactic nuclei candidate. QSO--Quasar.}
\end{deluxetable*}

\clearpage
\appendix

\section{SCUBA-2 Observation Archive} \label{app:obs}

Table\,\ref{tab:obs} presents the information for JCMT SCUBA-2 observations toward 70 High Latitude Planck Galactic Cold Clumps (HLPCs). We sort the observations by Galactic longitude and number the fields from 1 to 70. The serial number and the name of HLPC are listed in columns (1) and (2). The equatorial coordinates Right Accession (RA) and Declination (DEC) of the field center in Epoch J2000 are listed in columns (3) and (4). The project ID and scan pattern of the JCMT SCUBA-2 observation are listed in columns (5) and (6). The angular offset, which is defined by the angular distance from the field center to the center of the corresponding PGCC, is listed in column (7). The rms noise of the field is listed in column (8). As mentioned in Section\,\ref{data:distance}, the estimated distances and altitude are listed in columns (9)--(10). 

% Table 2: for the JCMT SCUBA-2 Observations
\startlongtable
\begin{deluxetable*}{cccccccccc}
% \rotate % rotate the wide table
% \centerwidetable
% \tabletypesize{\scriptsize}
% \tabletypesize{\small}
\tablewidth{0pt}
\linespread{1.1}
% \tablenum{1}
\tablenum{A1}
\tablecaption{Parameters of 70 High-latitude Planck Galactic Cold Clumps \label{tab:obs}}
\tablehead{
\colhead{No.} & \colhead{Field} & \colhead{RA} & \colhead{Dec.} & \colhead{Project ID} & \colhead{Scan Pattern$^{a}$} & \colhead{Offset} & \colhead{rms} & \colhead{Distance} & \colhead{Altitude} \\
\colhead{} & \colhead{} & \colhead{deg} & \colhead{deg} & \colhead{} & \colhead{} & \colhead{arcmin} & \colhead{\mjybeam} & \colhead{pc} & \colhead{pc}
}
\colnumbers
\startdata
1 & G4.13+35.75 & 238.3879 & -4.6406 & M14AU35 & Curvy Pong & 2.17 & 6.2 & 140 & 90 \\
2 & G4.17+36.67 & 237.6817 & -4.0717 & M15AI05 & CV Daisy & 0.55 & 13.8 & 130 & 87 \\
3 & G4.55+36.73 & 237.8058 & -3.7993 & M15AI05 & CV Daisy & 1.55 & 14.1 & 130 & 88 \\
4 & G4.80+37.02 & 237.7933 & -3.4799 & M15AI05 & CV Daisy & 4.53 & 13.7 & 130 & 88 \\
5 & G5.70+36.84 & 238.31 & -3.0125 & M16AL003 & CV Daisy & 0.01 & 15.9 & 120 & 83 \\
6 & G6.04+36.77 & 238.5362 & -2.8793 & M16AL003 & CV Daisy & 2.38 & 9.6 & 120 & 83 \\
7 & G27.31+37.33 & 246.8812 & 11.9261 & M15AI57 & Curvy Pong & 2.73 & 13.7 & 190 & 128 \\
8 & G37.52+44.57 & 242.6992 & 21.7625 & M15AI57 & Curvy Pong & 1.61 & 24.8 & 120 & 96 \\
9 & G45.12+61.11 & 225.65 & 29.3475 & M13AC22 & CV Daisy & 1.03 & 13.0 & 180 & 170 \\
10 & G45.16-36.19 & 320.2862 & -6.7184 & M15AI57 & Curvy Pong & 0.64 & 15.9 & 370 & -206 \\
11 & G48.63+34.42 & 256.0433 & 27.17 & M15AI57 & Curvy Pong & 1.08 & 14.0 & 220 & 133 \\
12 & G48.88+30.61 & 260.2037 & 26.2654 & M15AI57 & Curvy Pong & 0.57 & 10.2 & 240 & 134 \\
13 & G50.41-35.40 & 321.7487 & -2.6816 & M14AU02 & CV Daisy & 1.67 & 25.0 & 240 & -131 \\
14 & G53.44-36.25 & 323.7992 & -1.0448 & M15AI29 & CV Daisy & 0.6 & 5.7 & 260 & -143 \\
15 & G55.83-41.59 & 329.223 & -2.5211 & M19BP010 & CV Daisy & 2.3 & 21.3 & 170 & -105 \\
16 & G63.65+47.67 & 241.9708 & 40.0444 & M13AC22 & CV Daisy & 1.25 & 3.7 & 190 & 153 \\
17 & G92.49+42.88 & 242.3242 & 60.7558 & M13AC22 & CV Daisy & 0.81 & 15.0 & 240 & 176 \\
18 & G93.60+55.86 & 221.0292 & 54.3658 & M13AC22 & CV Daisy & 0.81 & 4.6 & - & - \\
19 & G96.05-50.34 & 355.8812 & 8.9585 & M15AI57 & Curvy Pong & 0.79 & 20.1 & 170 & -123 \\
20 & G105.07-38.06 & 357.7146 & 22.7272 & M16AL003 & CV Daisy & 2.32 & 18.3 & 240 & -141 \\
21 & G106.71-36.53 & 358.5827 & 24.566 & M16AL003 & CV Daisy & 1.61 & 20.0 & 190 & -105 \\
22 & G108.74-52.67 & 4.2204 & 9.2853 & M15AI57 & Curvy Pong & 2.6 & 21.4 & 190 & -144 \\
23 & G114.26-51.70 & 7.4162 & 10.7958 & M15AI57 & Curvy Pong & 0.87 & 26.3 & 210 & -151 \\
24 & G118.25-52.70 & 9.9721 & 10.0741 & M15AI57 & Curvy Pong & 0.6 & 25.4 & 190 & -144 \\
25 & G126.65-71.45 & 14.0508 & -8.5967 & M14AU02 & CV Daisy & 0.96 & 10.1 & 180 & -164 \\
26 & G128.30-69.65 & 14.7304 & -6.8572 & M15AI57 & Curvy Pong & 0.48 & 13.4 & 160 & -143 \\
27 & G128.76-69.46 & 14.9083 & -6.6946 & M15AI57 & Curvy Pong & 1.1 & 13.6 & 170 & -152 \\
28 & G128.94-46.39 & 17.1279 & 16.3175 & M16AL003 & CV Daisy & 3.56 & 32.1 & 190 & -130 \\
29 & G132.04-45.20 & 19.5372 & 17.2079 & M16AL003 & CV Daisy & 1.8 & 29.0 & 260 & -174 \\
30 & G133.71-46.64 & 20.5246 & 15.6283 & M19BP010 & CV Daisy & 2.17 & 25.8 & 180 & -123 \\
31 & G135.11-49.44 & 21.0113 & 12.685 & M19BP010 & CV Daisy & 4.59 & 24.2 & 230 & -165 \\
32 & G138.36-69.45 & 18.2708 & -7.2253 & M15AI57 & Curvy Pong & 0.74 & 17.5 & 160 & -143 \\
33 & G148.45+37.96 & 136.1146 & 65.9864 & M19BP010 & CV Daisy & 3.05 & 20.5 & 330 & 210 \\
34 & G152.54-47.36 & 32.7862 & 10.8591 & M14AU02 & CV Daisy & 0.96 & 8.8 & 180 & -125 \\
35 & G157.44+30.33 & 117.2383 & 59.6949 & M15AI29 & CV Daisy & 1.35 & 9.2 & 240 & 133 \\
36 & G158.51-33.99 & 43.7871 & 20.1953 & M16AL003 & CV Daisy & 0.0 & 14.0 & 220 & -112 \\
37 & G158.75-33.31 & 44.3617 & 20.6131 & M16AL003 & CV Daisy & 2.54 & 17.8 & 230 & -117 \\
38 & G158.86-34.19 & 43.9292 & 19.8755 & M16AL003 & CV Daisy & 0.01 & 15.0 & 240 & -127 \\
39 & G159.14-33.79 & 44.3671 & 20.0704 & M16AL003 & CV Daisy & 0.0 & 14.8 & 220 & -111 \\
40 & G159.21-34.28 & 44.1358 & 19.6369 & M15BI061 & CV Daisy & 0.0 & 19.2 & 240 & -128 \\
41 & G159.23-34.51 & 44.0096 & 19.4369 & M15BI061 & CV Daisy & 0.0 & 19.2 & 240 & -128 \\
42 & G159.41-34.37 & 44.1917 & 19.4608 & M15BI061 & CV Daisy & 1.95 & 13.8 & 210 & -106 \\
43 & G159.58-32.83 & 45.2717 & 20.6667 & M16AL003 & CV Daisy & 0.39 & 17.7 & 210 & -101 \\
44 & G159.66-34.31 & 44.4433 & 19.3998 & MJLSY14B & CV Daisy & 0.11 & 14.2 & 190 & -99 \\
45 & G161.43-35.60 & 44.9392 & 17.5759 & MJLSY14B & CV Daisy & 4.28 & 14.2 & 180 & -97 \\
46 & G161.67-35.92 & 44.8883 & 17.1405 & M16AL003 & CV Daisy & 0.01 & 15.9 & 180 & -97 \\
47 & G161.87-35.76 & 45.1096 & 17.1912 & M14AU02 & CV Daisy & 1.02 & 15.4 & 160 & -85 \\
48 & G174.35-39.98 & 50.0233 & 7.6149 & M14AU02 & CV Daisy & 1.06 & 11.2 & 150 & -89 \\
49 & G191.73-83.41 & 19.6517 & -24.5784 & M14AC02 & CV Daisy & 2.36 & 2.4 & 180 & -172 \\
50 & G197.98+33.10 & 128.4412 & 26.1953 & M14AU02 & CV Daisy & 0.44 & 17.2 & 260 & 151 \\
51 & G200.62+46.09 & 143.1058 & 27.3923 & M14AC02 & CV Daisy & 0.37 & 3.8 & - & - \\
52 & G203.57-30.09 & 71.9871 & -5.9299 & M16AL003 & CV Daisy & 0.0 & 12.5 & 220 & -99 \\
53 & G204.99+30.38 & 127.7183 & 19.6509 & M13BU09 & CV Daisy & 1.48 & 2.4 & 240 & 133 \\
54 & G210.90-36.55 & 68.7596 & -14.2279 & M15BI041 & CV Daisy & 2.37 & 5.1 & 150 & -77 \\
55 & G210.90+63.39 & 163.2525 & 24.9403 & M15AI57 & Curvy Pong & 1.75 & 15.2 & 110 & 108 \\
56 & G211.62+32.23 & 131.6992 & 15.1041 & M14AU02 & CV Daisy & 0.71 & 13.0 & 310 & 174 \\
57 & G220.55+60.12 & 161.2271 & 19.6588 & M14AU02 & CV Daisy & 0.95 & 12.9 & 150 & 144 \\
58 & G228.99+30.91 & 137.2925 & 1.3597 & M14AU15 & CV Daisy & 2.63 & 9.6 & 410 & 221 \\
59 & G235.60+38.28 & 146.4071 & 0.7633 & M15AI57 & Curvy Pong & 1.9 & 16.7 & 180 & 123 \\
60 & G235.67+38.00 & 146.2854 & 0.5485 & M15AI57 & Curvy Pong & 2.83 & 16.7 & 180 & 123 \\
61 & G240.03+68.75 & 173.2117 & 16.1356 & M15AI57 & Curvy Pong & 1.59 & 14.7 & 130 & 131 \\
62 & G251.29+73.32 & 179.3175 & 16.1378 & M15AI57 & Curvy Pong & 1.41 & 15.3 & 120 & 127 \\
63 & G259.63+31.89 & 155.3671 & -17.5 & MJLSY01 & Curvy Pong & 32.99 & 38.8 & 240 & 139 \\
64 & G343.12+58.61 & 212.5196 & 2.0519 & M14AU15 & CV Daisy & 2.32 & 9.3 & 140 & 127 \\
65 & G356.76+32.82 & 236.6396 & -11.2159 & M15AI57 & Curvy Pong & 1.88 & 29.4 & 150 & 89 \\
66 & G356.93+30.19 & 238.7129 & -12.8761 & M15AI57 & Curvy Pong & 0.98 & 28.8 & 180 & 102 \\
67 & G357.11+30.06 & 238.8917 & -12.8652 & M15AI57 & Curvy Pong & 0.7 & 26.7 & 160 & 92 \\
68 & G357.38+30.60 & 238.6437 & -12.3195 & M15AI57 & Curvy Pong & 0.51 & 25.0 & 150 & 84 \\
69 & G358.96+36.81 & 234.8908 & -7.1984 & M15AI57 & Curvy Pong & 1.73 & 27.5 & 130 & 88 \\
70 & G359.21+36.89 & 234.9154 & -6.9746 & M15AI57 & Curvy Pong & 1.4 & 14.2 & 120 & 83 \\
\enddata
\tablecomments{The serial number and the name of HLPC are listed in columns (1) and (2). The equatorial coordinates Right Accession (RA) and Declination (Dec.) of the field center in Epoch J2000 are listed in columns (3) and (4). The project ID and scan pattern of the JCMT SCUBA-2 observation are listed in columns (5) and (6). The angular offset, which is defined by the angular distance from the field center to the center of the corresponding PGCC, is listed in column (7). The rms noise of the field is listed in column (8). The distance derived from dust map is listed in column (9). The altitude from the Galactic mid-plane is listed in column (10). This table is available in its entirety in machine-readable form. }
\tablenotetext{a}{CV Daisy = Constant Velocity Daisy; Curvy Pong = Rotating Curvy Pong. }
\tablenotetext{b}{The rms noise within the ``cut-off radius''. }
\end{deluxetable*}

\section{Source Extraction} \label{app:extract}

To avoid large marginal noise features masquerading as sources, we set a ``cut-off radius'' within which we estimate noise and extract sources for each field. The ``cut-off radius'' depends on FoV. For the CV Daisy observation mode, the radius is set to 5\arcmin. For Curvy Pong, the radius is set 10\arcmin. One exception is the field 63 with FoV$\sim90\arcmin$, so we set the diameter to be $\sim67$\arcmin. We carefully select emission-free pixels and take the root mean square (rms) as a uniform noise $\sigma$ in each field (i.e., column (8) of Table\,\ref{tab:obs}). An intensity threshold of $3\sigma$, a step of $2\sigma$, and a minimum number of pixels (12 in our case) slightly larger than those contained in a JCMT beam are used for the input of the algorithm. In the output, ``leaves'' are the smallest structures and then defined as detected sources or sources hereafter. 

Figure\,\ref{fig:Extraction} displays all of the SCUBA-2 fields towards HLPCs. The ``cut-off radius'' utilized for source extraction is demarcated by the black dashed circles. The red contours demarcate the mask of extracted sources, while the outcomes of the 2D Gaussian fitting are visualized through orange ellipses. 

\begin{figure*}[!h]
\newcounter{1}
\setcounter{1}{\value{figure}}
\setcounter{figure}{0}
\renewcommand\thefigure{B\arabic{figure}}
\centering
\includegraphics[width=1.0\linewidth]{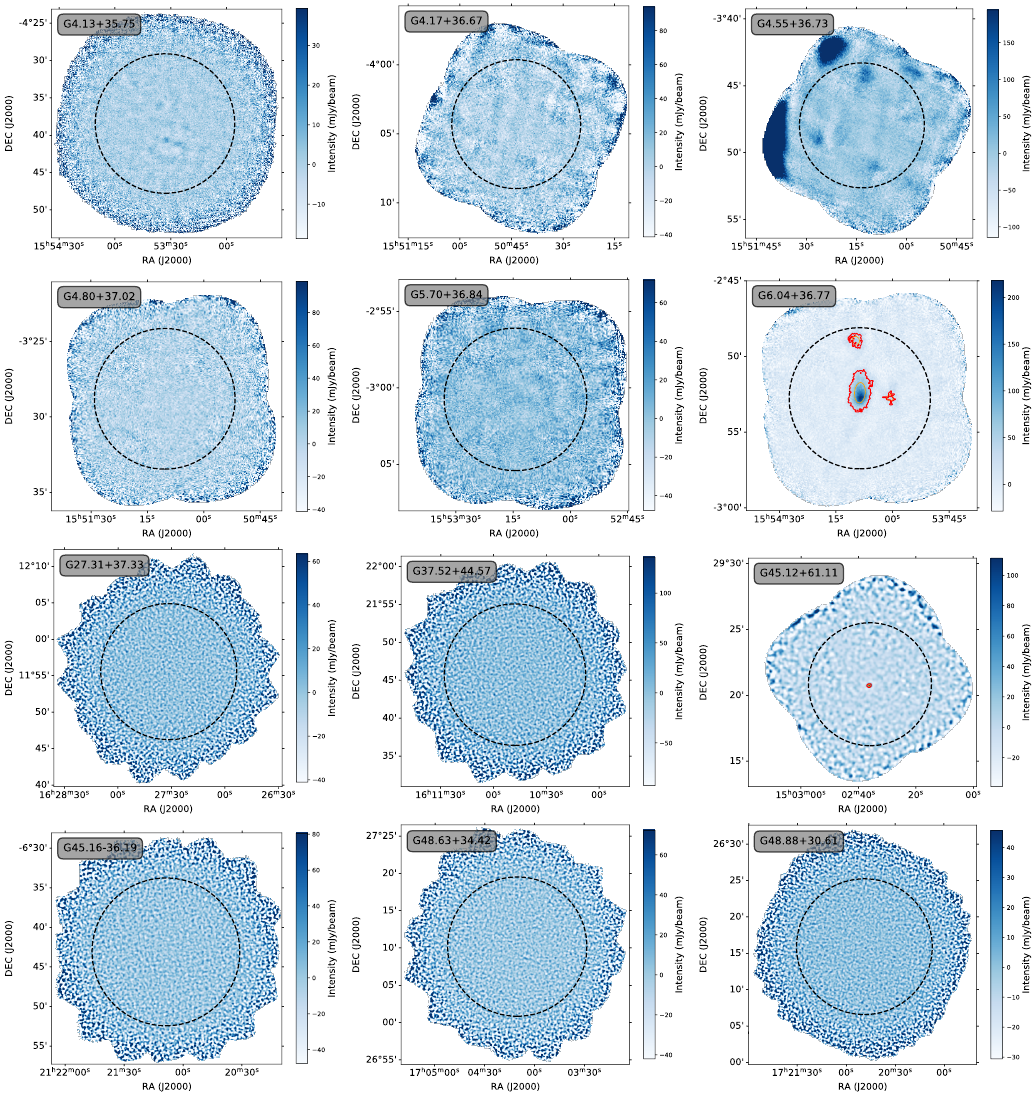}
\caption{Full atlas for 70 HLPCs in SCUBA-2 observations. The ``cut-off radius'' utilized for source extraction is demarcated by the black dashed circles. The red contours demarcate the mask of extracted sources, while the outcomes of the 2D Gaussian fitting are visualized through orange ellipses. The complete figure set (6 images) will be available in the online journal. \label{fig:Extraction}}
\end{figure*}

% \addtocounter{figure}{-1}
% \begin{figure*}[!h]
% \setcounter{1}{\value{figure}}
% \setcounter{figure}{0}
% \renewcommand\thefigure{B\arabic{figure}}
% \centering
% \includegraphics[width=1.0\linewidth]{HLP_atlas2.pdf}
% \caption{Continued.}
% \end{figure*}

% \addtocounter{figure}{-1}
% \begin{figure*}[!h]
% \setcounter{1}{\value{figure}}
% \setcounter{figure}{0}
% \renewcommand\thefigure{B\arabic{figure}}
% \centering
% \includegraphics[width=1.0\linewidth]{HLP_atlas3.pdf}
% \caption{Continued.}
% \end{figure*}

% \addtocounter{figure}{-1}
% \begin{figure*}[!h]
% \setcounter{1}{\value{figure}}
% \setcounter{figure}{0}
% \renewcommand\thefigure{B\arabic{figure}}
% \centering
% \includegraphics[width=1.0\linewidth]{HLP_atlas4.pdf}
% \caption{Continued.}
% \end{figure*}

% \addtocounter{figure}{-1}
% \begin{figure*}[!h]
% \setcounter{1}{\value{figure}}
% \setcounter{figure}{0}
% \renewcommand\thefigure{B\arabic{figure}}
% \centering
% \includegraphics[width=1.0\linewidth]{HLP_atlas5.pdf}
% \caption{Continued.}
% \end{figure*}

% \addtocounter{figure}{-1}
% \begin{figure*}[!h]
% \setcounter{1}{\value{figure}}
% \setcounter{figure}{0}
% \renewcommand\thefigure{B\arabic{figure}}
% \centering
% \includegraphics[width=1.0\linewidth]{HLP_atlas6.pdf}
% \caption{Continued.}
% \end{figure*}

\section{L183 Prestellar Cores} \label{app:psc}

Assuming that total emission $F_{\rm int}$ in column (6) of Table\,\ref{tab:det} is dust blackbody emission, then the mass of the three prestellar cores in L183 can be derived from,
\begin{equation}
    M = \frac{F_{\rm int} \mathcal{R} D^2}{\kappa_{\nu} B_{\nu}(T_{\rm dust})},
\end{equation}
where $D$ is distance of 120\,pc and $T_{\rm dust}$ is estimated from the temperature map which is derived from pixelwise SED fitting by \citet{Karoly2020L183}. As a result, $M_{1} = 1.8$\,\msun, $M_{2}=0.055$\,\msun, and $M_{3}=0.19$\,\msun. The mass of SMM1 is consistent with what has been derived in \citet{Karoly2020L183}, but the masses of SMM2 and SMM3 are much smaller. The reason is likely the 6 times better sensitivity in \citet{Karoly2020L183} than ours, resulting in more extended emission being included. 

The physical radius $R$ can be derived from deconvolved size by $\eta\sqrt{\sigma_{\rm maj}\sigma_{\rm min}} \times D$ where $\eta=2.4$ \citep{Rosolowsky2010BGPSII}. We obtain $R_{1}=7000$\,au, $R_{2}=2500$\,au, and $R_{3}=3300$\,au. And the averaged volume density of molecular hydrogen can be calculated assuming a sphere as, 
\begin{equation}
\bar n({\rm H_2}) = \frac{3M_i}{4\pi R_i^3\mu_{\rm H_2} m_{\rm H}}, \hspace{0.5cm}i\;=\;1,\;2,\;3.
\end{equation}
So we derive $\bar n_{1}=1.7\times10^5$\,cm$^{-3}$, $\bar n_{2}=1.0\times10^5$\,cm$^{-3}$, and $\bar n_{3}=1.6\times10^5$\,cm$^{-3}$, which are all consistent with values in \citet{Karoly2020L183}. 

\clearpage
\bibliographystyle{aasjournal}

      \bibliography{Scarcity}

\end{document}